\documentclass{PoS}
\usepackage{sidecap}

\title{Tests of Electric Polarizability on the Lattice}

\ShortTitle{Tests of Electric Polarizability on the Lattice}

\author{Victor X. Guerrero \\
Department of Physics, Baylor University, Waco, TX 76798-7316, USA \\
E-mail:\email{victor\_guerrero@baylor.edu}}
\author{Walter Wilcox \\
Department of Physics, Baylor University, Waco, TX 76798-7316, USA \\
E-mail:\email{walter\_wilcox@baylor.edu}}
\author{Joe Christensen \\
Department of Physics, Thomas More College, Crestview Hills, KY 41017, USA \\
E-mail:\email{joseph.christensen@thomasmore.edu}}

\date{November 24, 2008} 

\abstract{Using clover fermions on CP-PACS dynamical configurations, we consider a number of ways of measuring hadronic electric polarizability, an $|\mathbf{E}|^{2}$ effect in hadron masses, using lattice techniques. We consider the effects of periodic and Dirichlet boundary conditions, the field linearization postulate as well as a quantized electric field. We also consider two ways of formulating the classical vector potential which describes a uniform electric field in combination with the other possibilities.}

\FullConference{The XXVI International Symposium on Lattice Field Theory\\
		 July 14-19 2008\\
		 Williamsburg, Virginia, USA}

\begin{document}

\section{Introduction} 
Electric and magnetic polarizabilities characterize the rigidity of both charged and uncharged hadrons in external fields and are important fundamental properties of particles. In particular, the electric polarizability of a hadron characterizes the reaction of quarks to a weak external electric field and can be measured by experiment via Compton scattering. These sorts of quantities can be measured on the lattice\cite{1} using background fields for neutral particles. We report here on the ongoing work at Jefferson Lab by the \lq\lq polar'' collaboration to measure particle polarizabilities\cite{2,3,4}.

It turns out there are many different ways of introducing an electric field on the lattice. Here we discuss and present the results of various tests we have done in formulating a lattice quantum chromodynamics (QCD) calculation of neutral hadron electric polarizabilities. We shall consider the effects of periodic and fixed spatial boundary conditions, various types of quark sources and the field linearization postulate. We also consider two ways of formulating the classical vector potential for a uniform electric field from Maxwell's equations along with the other possibilities.

Hadron polarizability on the lattice can be measured as a change in mass in a uniform, external electric field, $\mathbf{E}$. The appropriate way of writing this is

\begin{equation}
\delta m = -\frac{1}{2} \alpha |\mathbf{E}|^2,
\end{equation}

\noindent where $\alpha$ is the electric polarizability coefficient. This form is necessary in order to give an invariant mass spectrum under the {\it continuous} Euclidean/Wick rotation, $\mathbf{E} \rightarrow e^{i\phi}\mathbf{E}$. On the lattice, one looks for a plateau in the effective mass shift plots vs. lattice time. Classical perturbation theory predicts that a weak electric field should induce a negative mass shift on the lowest mass quantum state from mixings with particles with higher mass but the same quantum numbers.
	 
We have used a small number of configurations (20) to test out various sources, boundary conditions and formulations. We are doing this in the context of the dynamical 2+1 CP-PACS clover configurations\cite{5}. We use a single mass, $\kappa_{ud} = 0.13580$ with $\kappa_s = 0.13640$ at $\beta =1.9$ on $20^3\times 40$ lattices ($a =0.098\pm .002$ fm). We include the electric field as a phase on the links which affect the Wilson term but not the clover loops. The values of the phases used (see Eqs.(2.1) and (2.2) below) are $\eta = 0.00036, -0.00072, 0.00144$, which can be combined to give two non-zero electric field values for $u$ and $d$ quark charges. This allows us to investigate the $|\mathbf{E}|^2$ behavior. The mass shifts used were symmetrized in the electric field direction,
\begin{equation}
\delta m(\mathbf{E}^2) \equiv \frac{1}{2}(\delta m (\mathbf{E}) + \delta m (-\mathbf{E})).
\end{equation}
In some of the tests we used a volume or spatial plane quark source to study the signal strength. Surprisingly, we found that this gave a good signal for the mesons but a poor or nonexistent one for the baryons. It turns out that when using an extended interpolation field for a hadron, the electric field phases will cancel on the sources for mesons but not for baryons. All of the results shown are therefore with point interpolation quark fields. Because of space limitations, we will show results only for the neutron mass shifts. (The neutral pion shifts are similar.)

\section{Classical $U(1)$ Gauge Invariance}

We start with the observation that two proposed ways of simulating a uniform electric field on the lattice (\lq\lq 4" is the lattice time direction and we are simulating a field in the \lq\lq 1" direction),
\begin{eqnarray}
U_{1}(n) \longrightarrow U_{1}(n)e^{i\eta n_4},\\
U_{4}(n) \longrightarrow U_{4}(n)e^{-i\eta n_1},
\end{eqnarray}
($\eta=eQEa^2$, $Q=\pm 1/3, \pm 2/3$, and $n_1$ and $n_4$ dimensionless) are connected in an exact \lq\lq classical" gauge invariant way in Wilson-type actions by the $U(1)$  transformations,
\begin{eqnarray}
U_{\mu}(n) &  \longrightarrow & G(n)U_{\mu}(n)G^{-1}(n+\hat{\mu}),\\
U^{\dagger}_{\mu}(n) & \longrightarrow & G(n+\hat{\mu})U_{\mu}(n)G^{-1}(n),\\
\psi(n) &  \longrightarrow & G(n)\psi(n), \\ 
\bar{\psi}(n)& \longrightarrow & \bar{\psi}(n)G^{-1}(n),
\end{eqnarray}
with the explicit form,
\begin{equation}
G(n) = e^{\pm i \eta n_1n_4},
\end{equation}
as long as one uses Dirichlet boundary conditions in the $n_1$ and $n_4$ directions for weak electric fields. (Dirichlet or uncoupled time edges are the usual way of simulating zero temperature.) Such non-periodic boundary conditions make a linear \lq\lq box'' in the 1-direction. One may also choose the field quantization condition,
\begin{equation}
\eta = \frac{2\pi}{N_1},
\end{equation}
where $N_1$ is the number of points in the 1-direction, in which case periodic boundary conditions may be used. For example, we may start with the links $U_1(n)e^{i\eta n_4}$ and $U_4(n)$. Choosing $G(n)=e^{i\eta n_1 n_4}$, this gives instead the links $U_1(n)$ and $U_4(n)e^{-i\eta n_1}$. These boundary conditions also allow an arbitrary shifting of the origin of the phases, obviating a problem seen in Ref.\cite{6}. 

\section{Data Interpretation}

\begin{figure} 
\begin{center}
\includegraphics[width=.8\textwidth]{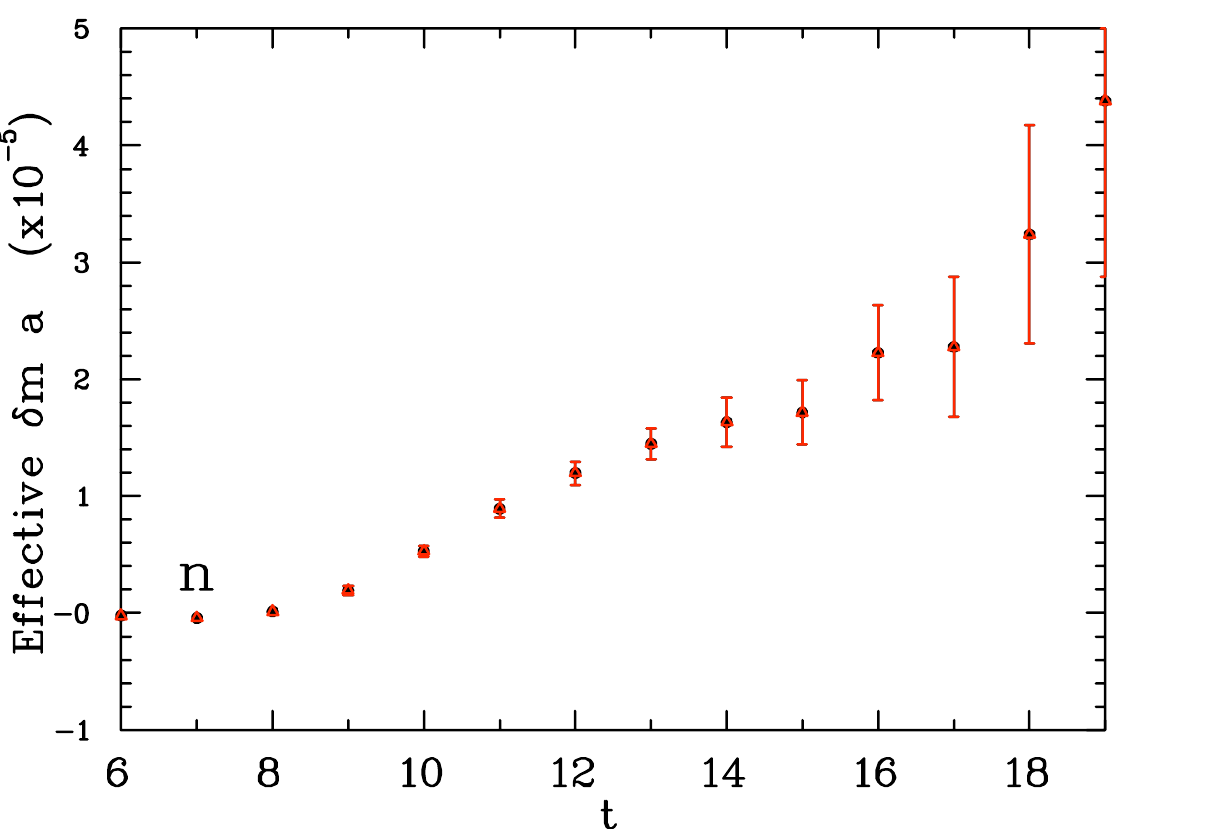} 
\caption{Point source results for effective mass shifts as a function of lattice time for
the smallest electric field for the neutron. The full exponential field was used with Dirichlet 
boundary conditions in the \lq\lq 1" and time directions. Results are from time dependent spatial links (black symbols) or spatially dependent
time links (red symbols). } 
\label{Figure 1} 
\end{center}
\end{figure} 

\begin{figure} 
\begin{center}
\includegraphics[width=.8\textwidth]{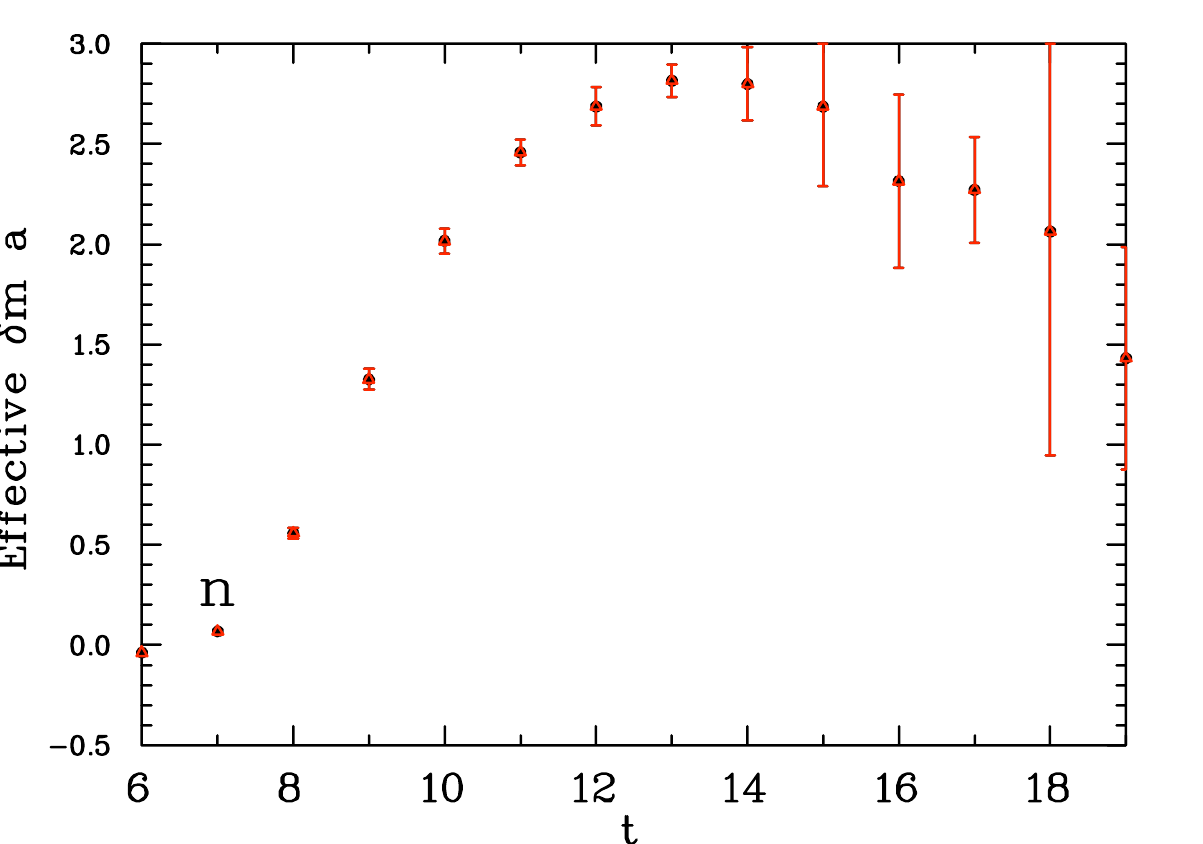} 
\caption{Point source effective mass shifts as a function of lattice time for
the smallest electric field for the neutron. The full exponential field was used. The quantization condition, Eq.(2.8), was used with periodic spatial boundary conditions and Dirichlet time boundary condition.
Results are from time dependent spatial links (black symbols) 
or spatially dependent time links (red symbols). } 
\label{Figure 2} 
\end{center}
\end{figure} 

Fig.~1 shows the results of a measurement of the effective mass shift on the neutron when the weak field electric field is introduced as a full exponential phase factor as in Eqs.(2.1) and (2.2). We are using Dirichlet boundary conditions in the electric field and time directions. Note that we see very good $|\mathbf{E}|^2$ behavior in the results. Although it is not clear we have a plateau in lattice time, the mass shift is seen to be positive, yielding a negative $\alpha$ from Eq.(1.2). We will comment on this in the Discussion section below. In addition, the two ways of introducing the electric field in Eqs.(2.1) and (2.2) are consistent with one another up to machine convergence.

Fig.~2 shows the results when an electric field, quantized as in Eq.(2.8), is introduced for the neutron. This time we use a periodic spatial boundary condition in the electric field direction and the full exponential phase. Nevertheless, the results with the two formulations in Eqs.(2.1) and (2.2) are still consistent. Again, a positive mass shift is seen. However, now the expected $|\mathbf{E}|^2$ behavior is badly broken between our lowest field and second electric field shifts. We see the results rise much too slowly to be consistent with $|\mathbf{E}|^2$.

\begin{figure}
\begin{center}
\includegraphics[width=.8\textwidth]{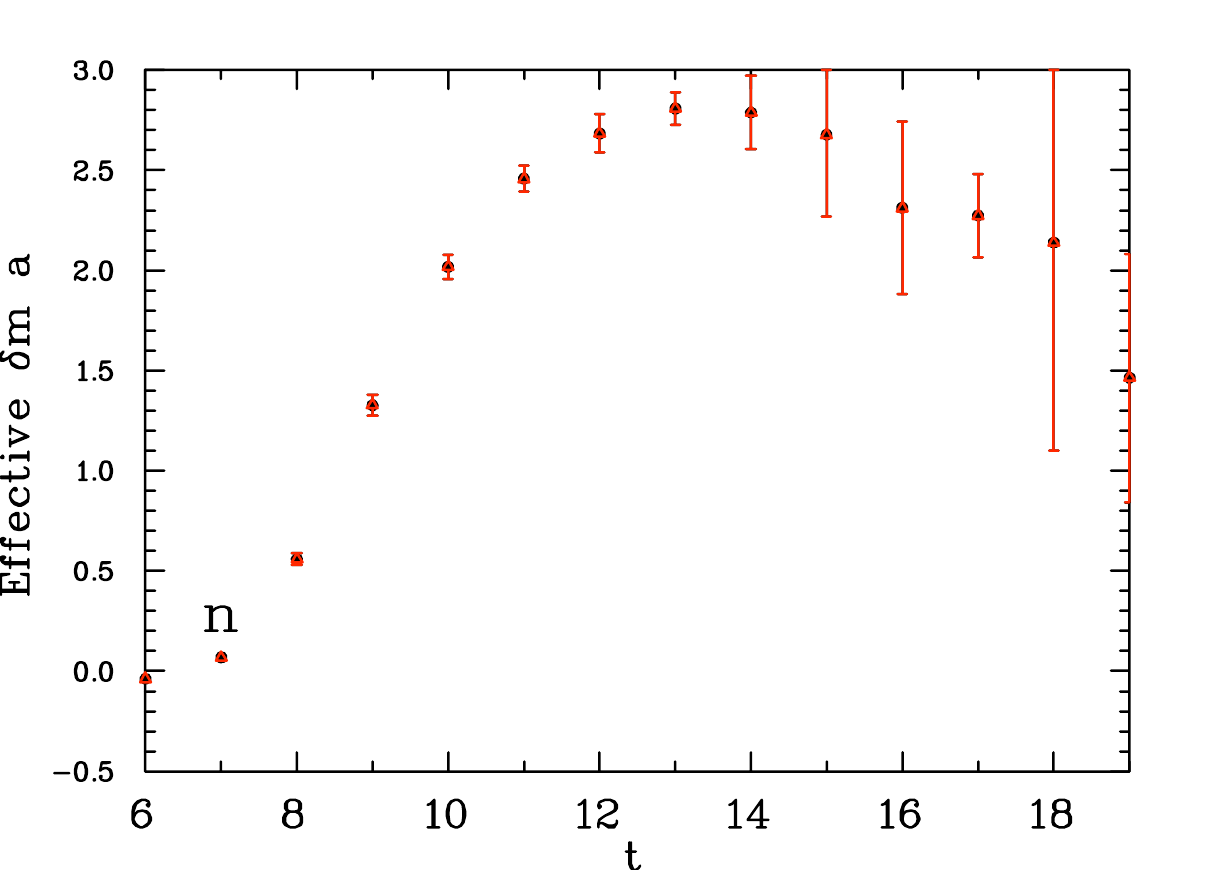} 
\caption{Point source effective mass shifts as a function of lattice time for
the smallest electric field for the neutron. The full exponential field was used. The quantization condition, Eq.(2.8), was used with Dirichlet boundary conditions in the \lq\lq 1" and time directions. The data are similar, but not identical, to Fig.~2. Results are from time dependent spatial links (black symbols) or spatially dependent time links (red symbols). } 
\label{Figure 3} 
\vspace{1cm}
\includegraphics[width=.8\textwidth]{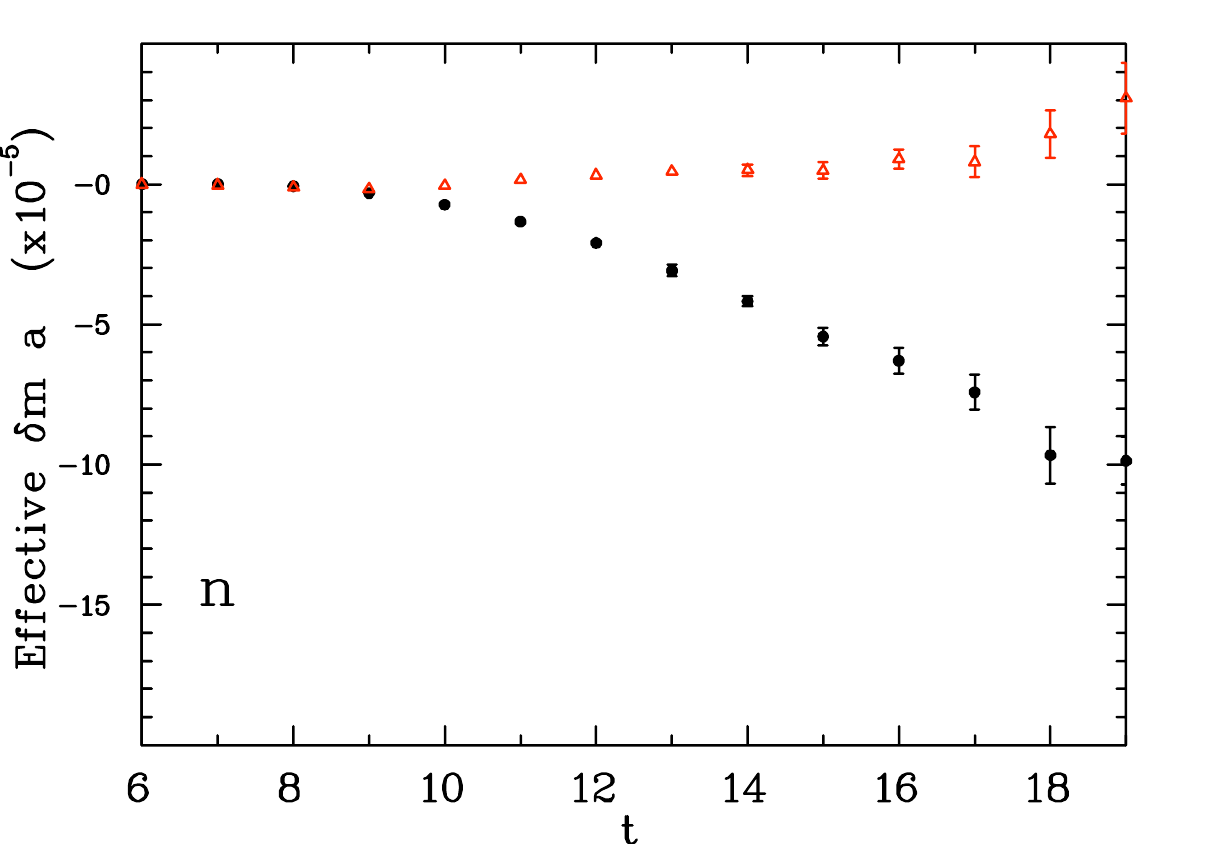} 
\caption{Point source effective mass shifts as a function of lattice time for
the smallest electric field for the neutron. The electric fields were linearized according to
Eqs.(3.1) and (3.2) with Dirichlet boundary conditions in the \lq\lq 1" and time directions. Results are from time dependent spatial links (black symbols) or spatially dependent time links (red symbols).} 
\label{Figure 4} 
\end{center}
\end{figure} 

Fig.~3 is much like Fig.~2, but uses Dirichlet boundary conditions in the electric field direction instead of periodic boundary conditions. Although the numerical results are very similar to Fig.~2, careful comparison shows that Figs.~2 and 3 are slightly different, as they should be. Nevertheless, the two ways of putting in the electric field, Eqs.(2.1) and (2.2), agree very well with one another. This measurement also suffers from non-$|\mathbf{E}|^2$ behavior. Again, we will comment below on the positive mass shifts.

Finally, we examine the effect of the field linearization postulate in Fig.~4. By this we mean the replacement of the phase 
\begin{equation}
e^{i\eta n_4} \longrightarrow (1 + i\eta n_4),
\end{equation}
in the $U_1$ link case and 
\begin{equation}
e^{-i\eta n_1}\longrightarrow (1 - i\eta n_1),
\end{equation} 
in the $U_4$ link case. Everything else in the calculation was kept the same as in Fig.~1. Even though the linear terms were kept quite small, one can see that the results are very different from before and that the gauge invariance seen in the other figures is lost. We verified that these small field mass shifts are, however, proportional to $|\mathbf{E}|^2$. The mass shift for the electric field put in as a $n_4$ dependent $U_1$ link is negative, consistent with our previous determinations. The other formulation using a $n_1$ dependent $U_4$ ink yields a much smaller mass shift.

\section{Discussion}

Superficially, the gauge invariance seen in Figs.~1, 2 and 3 would seem encouraging. However, classically, we would expect a decreased rather than increased mass when the electric field is introduced. In order to understand this effect, let us consider expanding the electric field $U(1)$ phase to second order in the action. We have
\begin{equation}
e^{iaqA_{\mu}} \longrightarrow (1 + iaqA_{\mu} - \frac{1}{2}(aqA_{\mu})^2).
\end{equation}
Higher order terms will not contribute in the present case since we are examining an $|\mathbf{E}|^2$ effect. In Wilson-type actions, one can show that the linear $A_{\mu}$ term will couple to the exactly conserved lattice vector current, $J_{\mu}^{latt}$. However, the $A_{\mu}^2$ term has no continuum action analog; it is the classical analog of a \lq\lq tadpole" term in the gauge action\cite{7}. Inclusion of this term in the lattice simulations will couple to the original gauge action, but with the gauge field link $U_1$ or $U_4$ replaced with $U_1(1-\frac{1}{2}(\eta n_4)^2 )$ or $U_4(1 -\frac{1}{2}(\eta n_1)^2)$, respectively. Thus, the links associated with the phases are effectively reduced in magnitude. This means that all Green functions which use these links will experience a reduced magnitude as a function of lattice time. When compared to the zero field case, this effect will simulate a completely spurious increased mass\cite{8}.

The obvious way of avoiding the tadpole term is to linearize the phases. The linear field results in Fig.~4 show, unfortunately, that the casualty is classical gauge invariance. The spatial link result using Eq.(3.1) gives a negative effective mass shift as a function of lattice time, similar to our previous measurements\cite{3}. We also see that the time link result using Eq.(3.2) is inconsistent with the spatial link measurement. However, these two ways of putting in the linear field in the context of the 4-point polarizability function\cite{9} are distinguished by the lattice physics involved. One may show that the position operator on the lattice introduces higher momentum states\cite{10}. On the other hand, the linear representation in Eq.(3.1) is an exact zero momentum object on the lattice consistent with the desired measurement.

\section{Conclusions}
We have displayed an exact classical gauge invariance which connects two proposed exponential ways of formulating a uniform electric field and examined the effects of the fields in a lattice simulation of electric polarizability. In the case of weak electric fields and Dirichlet boundary conditions, we have verified the gauge invariance and seen positive neutron mass shifts. We have also examined the effect of using a quantized electric field and seen that the results are classically gauge invariant for periodic or \lq\lq 1" direction Dirichlet boundary conditions. However, we have seen that the quantized results violate the weak field limit necessary to obtain the correct $|\mathbf{E}|^2$ behavior for electric polarizability. Moreover, we have argued that the positive mass shifts seen with the full exponential representation have contributions from an incorrect classical tadpole term in the electromagnetic $U(1)$ field. The linearized $U(1)$ links are free of the spurious tadpole term, but unfortunately are no longer classically gauge invariant. Nevertheless, the representation of the constant electric field using linearized $U_1$ links as in Eq.(3.1) results in a correct reduction in the effective neutron mass and also correctly avoids introducing non-zero momenta in the lattice measurement of electric polarizability.

\section{Acknowledgements}
This work is supported in part by U.S. Department of Energy under grant DE-FG02-95ER40907. Our calculations were done with QDP++ and Chroma software at Jefferson Lab. We gratefully acknowledge the use of the CP-PACS configurations as well as a grant from the Baylor University Research Committee. We also thank F. X. Lee for helpful discussions and calculations.


\begin{thebibliography}{00}
\bibitem{8} This term can be partially cancelled by a mean field strategy similar to Ref.\cite{7}; this has yet to be studied numerically.
\end{thebibliography}
\end{document}